\documentclass[10pt]{iopart}
\begin{document}

\title{Ballistic transport is dissipative: the why and how}

\author{Mukunda P. Das\footnote[3]
{To whom correspondence should be addressed
(mukunda.das@anu.edu.au)}}
\address{Department of Theoretical Physics,
Institute of Advanced Studies,\\
The Australian National University,
Canberra, ACT 0200, Australia}

\author{Frederick Green}
\address{Department of Theoretical Physics,
Institute of Advanced Studies,\\
The Australian National University,
Canberra, ACT 0200, Australia\\
and\\
School of Physics, The University of New South Wales
Sydney, NSW 2052, Australia}

\begin{abstract}
In the ballistic limit, the Landauer conductance steps
of a mesoscopic quantum wire have been explained by
coherent and dissipationless transmission of individual
electrons across a one-dimensional barrier.
This leaves untouched the central issue of conduction:
a quantum wire, albeit ballistic, has finite resistance
and so must dissipate energy. Exactly {\em how} does the
quantum wire shed its excess electrical energy? We show that the
answer is provided, uniquely, by many-body quantum kinetics.
Not only does this inevitably lead to universal quantization
of the conductance, in spite of dissipation; it fully resolves
a baffling experimental result in quantum-point-contact noise.
The underlying physics rests crucially upon the action of
the conservation laws in these open metallic systems.
\end{abstract}

In metallic transport there is at least one inviolate universal:
where there is finite resistance, there must be finite
power dissipation. That rule (otherwise known as the
fluctuation-dissipation theorem) applies as much to
mesoscopic {\em ballistic} conduction as to any other type.
Despite this, the physical meaning of mesoscopic dissipation
in a ballistic wire, or quantum point contact (QPC),
persists as the subject of theoretical speculation.
Ascribing these inevitable energy losses to unspecified
processes, deep in the QPC's leads
\cite{p2},
may provide an insight. It does not solve the physics. 

The vexed issue of ballistic dissipation has been put
squarely on the table by, among others, Davies
\cite{davies}
and Agra\"{\i}t {\em et al}.
\cite{agrait}.
We aim to shed further light on this topic. We do so firmly
within the established canon of quantum kinetics
\cite{mahan}.

Contemporary successors to Landauer's transport theory
\cite{p2,p1,p3}
have afforded a fresh understanding of mesoscopic transport.
One is dealing with small, possibly near-molecular, structures.
Consequently they experience an unprecedented level of openness
to their macroscopic environment. New kinds of
understanding -- new kinds of physics -- have been perceived
as essential to progress in this realm
\cite{p2}.

There are now two striking physical signatures of transport in
quantum point contacts:
(a) discretization of conductance into its classic
``Landauer steps'', in units of $2e^2/h$, and
(b) the totally unexpected peak structures in the noise
of a QPC driven at constant current
\cite{rez}.
These properties will be our points of reference.

Coherent transmission of an electron wave
function offers a ready and plausible explanation of
one-dimensional conductance quantization
\cite{p2,p1}.
Nonetheless it is recognized
\cite{davies,agrait,frensley}
that, within single-particle quantum mechanics,
there is no answer to the simple question:
{\em where is the dissipation in a ballistic quantum point contact}?

The question, as posed, is far more than academic.
Nano-electronics is almost with us. Effective and trustworthy
nano-electronic design will demand more than plausible models.
It will demand a quantitative accounting of every dominant physical process.
Not the least of these is dissipation, near equilibrium and well beyond.

Let us restate plainly the issue at the heart of all metallic conduction.
Any finite conductance $G$ must dissipate electrical energy at the rate
$P = IV = GV^2$, where $I = GV$ is the current and $V$ the potential
difference across the terminals of the driven conductor.
It follows that there is always some explicit mechanism
(impurity-mediated electron-hole excitations,
optical phonon emission, etc.) by which the energy gained
by carriers from the source-drain field is passed
incoherently to the surroundings.

Alongside any elastic and coherent scattering processes,
inelastic processes must always be in place. Harnessed together,
they fix $G$. Yet it is only the energy-dissipating mechanisms
that secure the {\em thermodynamic stability} of steady-state conduction.

The microscopic understanding of the universal power-loss formula
$P = GV^2$ has been with us for some time
\cite{kubo}--\cite{wims}.
It resides in the fluctuation-dissipation theorem (FDT),
valid for every resistive device, at every scale.
The theorem precisely expresses the requirement of
thermodynamic stability. From it comes the conclusion that
\cite{DG1,DG2}

(i) inelasticity is necessary and sufficient to stabilize current
flow at finite $G$;

(ii) ballistic quantum point contacts have finite
$G \propto e^2/\pi\hbar$; therefore

(iii) the physics of energy loss is indispensable
to any theory of ballistic transport.

\noindent
All of the physics of inelastic scattering is lost if
the ballistic conductance is modelled by coherent one-body
collisions alone. Inelasticity and dissipation emerge only from
the higher-order, yet dominant, electron-hole pair correlations
(the ``vertex corrections'') in potential scattering.
These never appear at the one-body level
\cite{mahan}.

Coherence entails elasticity. Elastic scattering is always
loss-free; it conserves the energy of the scattered particle.
Energy loss requires many-body processes. Many-body scattering
cannot be described as single-carrier transmission.

To incorporate the physics of energy dissipation, underpinning
the microscopic description of ballistic transport,
we recall that open-boundary conditions imply the intimate
coupling of the QPC channel to its interfaces with the reservoirs.
The interface regions must be treated as an integral part of
the device model. They are the actual sites for strong scattering
effects: {\em dissipative} many-body events
as the current  enters and leaves the ballistic channel,
and {\em elastic} one-body events as the carriers interact
with background impurities, the potential barriers that
confine and funnel the current, and so forth. 

It is essential to subsume the interfaces within
the total kinetic description of the ballistic channel.
For an electrically open conductor, the canonical requirement
of strict charge conservation is satisfied uniquely by the
explicit supply and removal of current via an external generator
\cite{sols}.
Immediately, it follows that the current cannot at all be
determined by the local conditions in the reservoirs.
This fundamental result sets the quantum kinetic approach
entirely apart from alternative treatments
\cite{p2},
which assume that a current depends directly
on the differences of physical density between
reservoirs separately attached to the channel.

Earlier
\cite{DG1}
we applied microscopic response theory to
obtain the conductance of an open sub-band in a
one-dimensional ballistic wire.
There, the effective mean free paths
$\lambda_{\rm el}$ and $\lambda_{\rm in}$ for
both elastic {\em and} inelastic scattering are no longer delimited
by their bulk values for the wire material; the path lengths become
constrained by the presence of current injection and extraction,
located in the leads connected to the wire and separated by $L$. The
total effective mean free path is given by
$\lambda_{\rm tot}^{-1} = \lambda_{\rm in}^{-1} + \lambda_{\rm el}^{-1}$.

In the optimum ballistic limit, both of the elementary constituent
paths are equal to $L$. A little algebra leads to $G = 2e^2/h$:
the Landauer conductance of a single, one-dimensional, ideal channel
\cite{DG1}.
We draw attention to the fact that none of the supplemental
assumptions, essential to deriving conductance quantization
from the coherent-transmission theories
\cite{p2,p3},
is in the least necessary
\cite{DG1}.
Indeed the Landauer formula (LF) emerges straight
from the standard quantum kinetics of a driven system
\cite{mahan},\cite{kubo}--\cite{wims}.

Most important to the LF, as a microscopic entity,
is the manifest and central role of inelastic energy loss,
through $\lambda_{\rm in}$. It summarizes the energy-absorbing
role of the electron-hole {\em fluctuations}. In their
explicitly dissipative action at this fundamental level,
they embody the FDT, one of the pivots for quantum transport.
The other pivot is charge conservation
\cite{sols,wims}.
The {\em explicit} roles of energy loss and open-system
charge conservation cannot be dispensed with.

The FDT states that, without dissipation,
there is no resistance in a quantum wire
\cite{depic}.
Granting that, one has to conclude that the overall,
finite, Landauer conductance -- as we observe it in the
ballistic regime -- ``belongs''
as much to the leads (which are decidedly non-ballistic)
as to the contact in between. The channel {\em and} the leads are one
whole, integrated, mesoscopic assembly: the ``ballistic wire''.

As far as conductance is concerned, we have traced out how
quantum kinetics {\em per se}, in tandem with the charge-conserving
and dissipative physics at the open boundaries
\cite{sols,wims},
demands the presence and efficacy of inelastic scattering.
This extends to an ideal ballistic channel. Inelastic and elastic
effects always co-exist
\cite{mahan}.

Microscopically, the concept of a specific, direct and active
physical role for the dissipative boundary leads cannot be separated
from that of an even perfectly ballistic wire. We may ask: what other
evidence exists for this {\em intrinsically kinetic} understanding
of conductance? It is the non-equilibrium noise of a QPC. 

Careful measurements of non-equilibrium noise in a quantum wire
have been published by Reznikov and colleagues
\cite{rez}.
Alongside the more or less anticipated noise spectrum of
a quantum point contact at fixed voltage, there is an additional
very puzzling result. Although accepted models
\cite{p3}
fail to predict any fine structure whatsoever in the noise of
the wire driven at constant channel {\em current}, the data exhibit
a systematic set of well-defined, strong peaks at the gate-bias point
where the carrier density accesses the first conduction sub-band.

The noise data for constant channel current remained
without a satisfactory theoretical explanation until recently.
There is now a microscopically based account for the
Reznikov {\em et al.} measurements
within the same strictly conserving kinetics that serves
for the conductance of the LF.
We cite reference
\cite{DG3},
covering the theory of, and full quantitative comparison
with, the complete noise data of reference
\cite{rez}:
constant-voltage as well as the (formerly) baffling
constant-current sets.

Three conditions should be met by any theory
for the results of Reznikov {\em et al}.
\cite{rez}
and any comparable experiment.
\smallskip

\noindent
(a) Both at constant voltage and constant current, the
experimental conditions take the quantum wire well outside
the weak-field regime
\cite{DG3}.
Thus, strictly linear models ought not to be used
except with great caution.
\smallskip

\noindent
(b) A microscopic description of the fluctuations will apply
{\em if and only if} it incorporates the same dissipation physics
which, as we now know, is essential to fixing $G$.
\smallskip

\noindent
(c) One should expect a formal harmony
between the analysis of conductance, and that of the noise fluctuations
(at weak field, the latter encompass the former via the FDT).
In particular, a unified model will automatically satisfy the
conservation relations that exist
over and above the fluctuation-dissipation theorem.
\smallskip

A less familiar and no less fundamental conservation relation
concerns the compressibility of an electron fluid in a
conductive channel
\cite{pinoz}.
This relation has an immediate and explicit link with
non-equilibrium noise behaviour in a QPC
\cite{DG3,csr}.
The outworking of the non-equilibrium compressibility rule is,
quite directly, the emergence of the peak structures
observed in the excess noise of a quantum point contact
\cite{DG3}.
This striking instance offers a window onto the
central function of conservation laws in the physics
of transport at mesoscopic dimensions and below.

In summary the orthodox kinetic analysis of transport
\cite{mahan},\cite{kubo}--\cite{DG2}
suffices, uniquely, for a detailed and innately microscopic account of
mesoscopic conductance and non-equilibrium noise.
Specifically this is
because the manifest role of inelastic
dissipation, alongside elastic scattering, is
accorded its full and true physical importance.
The description is perfectly conserving and seamless.

Quantum kinetics resolves, in a wholly natural way,
the long-mooted ``problem'' of ballistic dissipation. It accurately
reproduces the current response of a mesoscopic conductor -- its
Landauer conductance -- free of {\em ad hoc} assumptions.
It does as much for the associated current noise.
The keys to the canonical understanding of transport are open-system charge
conservation and the physical reality of dissipative scattering.

\end{document}